\documentclass[]{article}
\usepackage{amssymb,amsmath}
\def\d{\partial}

\def\Bf#1{\mbox{\boldmath $#1$}}

\def\Im{{\rm Im}}

\def\2{{1\over 2}}
\def\N2{${\cal N}=2$}
\def\4N{${\cal N}=4$}
\def\1N{${\cal N}=1$}
\def\F{{\cal F}}
\def\f{{\sf F}}
\def\C{{\sf C}}
\def\sign{{\rm sign}}

\begin{document}

\begin{flushright}
FIAN/TD-07/02\\
ITEP/TH-25/02\\
EMPG-02-12
\end{flushright}
\vspace{0.5cm}

\begin{center}
{\LARGE \bf WDVV Equations as Functional Relations }
\end{center}
\vspace{0.5cm}

\begin{center}
{\Large H.W. Braden\footnote{Department of Mathematics and Statistics,
University of Edinburgh, Edinburgh EH9 3JZ Scotland;
e-mail address: hwb@ed.ac.uk }} {\large and} {\Large
A. Marshakov\footnote{Theory
Department, Lebedev Physics Institute, Moscow
~117924, Russia and ITEP,
Moscow 117259, Russia; e-mail address: mars@lpi.ru;
andrei@heron.itep.ru}}
\\
\end{center}

\begin{quotation}
\noindent
We discuss the associativity or WDVV equations and demonstrate that
they can be rewritten as certain functional relations between the {\it second}
derivatives of a single function, similar to the dispersionless Hirota
equations. The properties of these functional relations are further
discussed.
\end{quotation}

\section{Introduction}

The associativity or Witten-Dijkgraaf-Verlinde-Verlinde (WDVV) equations
\cite{WDVV} have been widely discussed in connection with various problems of
mathematical physics for over ten years. They have arisen in the context of
quantum cohomologies and mirror symmetry, and also with regard to
multidimensional supersymmetric gauge field theories.
In their most general form these equations may be expressed as \cite{MMM}
\begin{equation}
\label{WDVV}
\f_i\cdot \f_j^{-1}\cdot\f_k = \f_k\cdot \f_j^{-1}\cdot\f_i
\qquad \forall i,j,k.
\end{equation}
Here the matrices $\|\f_i\|_{jk} \equiv \F_{ijk}$ are constructed from the
third derivatives of the function $\F({\bf a})$,
\begin{equation}
\label{thirdder}
\F_{ijk} =\dfrac{\d^3\F}{\d a\sp{i}\,\d a\sp{j}\,\d a\sp{k}}.
\end{equation}
When $\F$ is a function of one or two variables then (\ref{WDVV}) are empty:
that is they are satisfied by any function. For more variables however,
despite the simplicity of their compact matrix form, these equations
form a highly nontrivial overdetermined system of nonlinear partial
differential equations satisfied by the function $\F$.

In their original two-dimensional topological field theoretic setting
the WDVV equations (\ref{WDVV}) were supplemented by a further constraint:
there was a distinguished coordinate such that the matrix
$\F_{1jk}$ was a constant.
This came from the existence of a distinguished operator, the identity
operator $\Phi_1$, and the third derivatives of $\F$ being related to
three-point functions via
$\langle \Phi_i\Phi_j\Phi_k\rangle=  \F_{ijk}$.
This distinguished coordinate meant their was a preferred ``metric'', $\f_1$.
Associated with this class of solutions to (\ref{WDVV}) -- together with
a quasi-homogeneity condition -- Dubrovin introduced the notion of
Frobenius manifold \cite{Dub}. However, there are physically
interesting solutions of (\ref{WDVV}) that fail to possess these extra
properties and for these the notion of Frobenius manifold fails to
encapsulate their geometry.

This more general class of solutions to (\ref{WDVV}) includes
some prepotentials arising from
low-energy effective actions of \N2 supersymmetric gauge theories in four
dimensions via Seiberg-Witten theory.
Supersymmetry typically restricts the possible {\em geometries} encountered in
field theories. Thus sigma models have target spaces that
are Riemannian, or possibly K\"ahler or hyper-K\"ahler depending on
the number of supersymmetry generators.
For \N2 SUSY gauge theories the moduli space of vector multiplets
is a special\footnote{
In the original supergravity context such a manifold was called
``rigid special K\"ahler" with ``special K\"ahler" referring to the
local (supergravity) setting; since \cite{Freed} the local setting is
often known by ``projective special K\"ahler". See \cite{cortes}
for a survey of these manifolds.}
K\"ahler manifold \cite{speKa,SW}.
This means that on the moduli space there exists a single {\em holomorphic}
function $\F ({\bf a})$ whose second derivatives
\begin{equation}
\label{pematr}
T_{ij} =\dfrac{\d^2\F}{\d a\sp{i}\,\d a{\sp{j}}}
\end{equation}
determines the metric via
\begin{equation}
\label{metric}
ds\sp2= g_{i{\bar j}}\,d a\sp{i} d\overline{ a\sp{j}},\qquad
g_{i{\bar j}}=\dfrac{1}{2}(\Im T)_{ij}.
\end{equation}
Here $a\sp{i}$ are complex coordinates on the  K\"ahler manifold
(and $\overline{ a\sp{j}}$ their complex conjugates). The connection
with Seiberg-Witten theory is that the second derivatives $T_{ij}$
coincide with the matrix elements of the period matrix of the
Seiberg-Witten auxiliary curve and $\F$ is the prepotential.
In \cite{GKMMM} it was argued that this auxiliary curve could be identified
with the spectral curve of a completely integrable system. Indeed, the phase
space of (appropriate algebraically) completely integrable systems can
be identified as a toric fibration (the angles $x\sp{i}$) over a special
K\"ahler manifold (the actions $y_j$). The K\"ahler form is then
\begin{equation}
\label{kform}
\omega=dx\sp{i}\wedge dy_i
=\dfrac{\sqrt{-1}}{2}\, (\Im T)_{ij}\, d a\sp{i}\wedge d\overline{ a\sp{j}},
\end{equation}
and the real coordinates $\{x\sp{i},y_j\}$ are related to complex coordinates by
\begin{equation*}
\dfrac{\d }{\d a\sp{i}}=\dfrac{1}{2}\bigg(\dfrac{\d }{\d x\sp{i}}
-T_{ij}\, \dfrac{\d }{\d y_{j}}\bigg).
\end{equation*}
The K\"ahler potential here is
$K({\bf a}, {\bf\bar a})=\frac{1}{2}\bigg(\overline{ a\sp{i}}\,\dfrac{\d \F}
{\d a\sp{i}}\bigg)$.
A special K\"ahler manifold  has two natural connections associated with it.
There is the Levi-Civita connection $D$, with Christoffel symbols
$\Gamma_{{\bar i}jk}=-\frac{\sqrt{-1}}{4}\F_{ijk}$, and there is also
a flat torsionfree connection $\nabla= D+A$ such that $\nabla d x\sp{i}=0
=\nabla d y_{j}$.
Here the connection $A$,
\begin{equation}
\label{connection}
A\bigg(\dfrac{\d }{\d a\sp{i}}\bigg)=\big(A_j\big)\sp{\bar l}_{\ i}\,
d a\sp{j}\otimes \dfrac{\d }{\d \overline{ a\sp{l}}}
=\dfrac{\sqrt{-1}}{4}\, \bigg(g\sp{{\bar l}m} \F_{jim}\bigg)\,
d a\sp{j}\otimes \dfrac{\d }{\d \overline{ a\sp{l}}},
\end{equation}
satisfies $0=\d\sp{D} A=\d A+\Gamma\wedge A-A\wedge \Gamma$.
Because the Levi-Civita connection connection is torsionfree we may also
write this as $D_i A\sp{\bar l}_{\, jk}=D_j A\sp{\bar l}_{\, ik}$
with $A\sp{\bar l}_{\, jk}=(A_j)\sp{\bar l}_{\ k}$.
We refer to \cite{mars, brkr} for further material on the connections
between Seiberg-Witten theory and integrable systems.

Following the work of \cite{GKMMM}, it seemed to be very important that the
function $\F$ satisfied some well-known nonlinear integrable differential
equations.
The connection between Seiberg-Witten theory and the WDVV equations is
that {\it some} prepotentials lead to solutions of (\ref{WDVV}). Certainly not
all prepotentials yield solutions and a characterisation of what
special K\"ahler manifolds give solutions is still sought.
Now the geometric origin of the WDVV equations in Seiberg-Witten theory
appears to be completely different from that of Frobenius manifolds.
Seiberg-Witten theory lacks the analogue of the identity operator
and constant ``topological metric''. Indeed there appears no obvious connection
between the (nonholomorphic) metric of special K\"ahler geometry,
built out of second derivatives of the prepotential, and the
``topological metric'' $\f_1$ of the Frobenius manifolds which is a third
derivative.
Further,  the WDVV equations were shown to be covariant with respect to
arbitrary symplectomorphisms in \cite{dWM}, which is quite natural in
the special K\"ahler setting, but such do {\em not} preserve the
Frobenius manifold structure.

The geometric origin underlying the WDVV equations has yet be clarified.
In this note we will reformulate these equations as functional relations.
In particular, they will be in terms of second derivatives.
These functional relations are reminiscent of the Hirota equations
and we will give an example making this clearer.

\section{WDVV and dispersionless Hirota equations}

The WDVV equations usually follow from the crossing relations
\begin{equation}
\label{crossing}
\sum_k C^k_{\ ij}C^n_{\ kl} = \sum_k C^k_{\ il}C^n_{\ kj}
\end{equation}
for the structure constants of some algebra
\begin{equation}
\label{alg}
\phi_i\circ\phi_j =\sum_k  C^k_{\ ij}\phi_k.
\end{equation}
It is useful also to write (\ref{crossing}) in the matrix form
\begin{equation}
\label{crossmatr}
\C_i\cdot \C_j = \C_j\cdot \C_i,\qquad \forall i,j
\qquad  (\C_i)\sp{l}_{\ m} = C_{\ im}\sp{l}.
\end{equation}
Equations (\ref{crossing}) are algebraic relations and they turn into the
WDVV system of nonlinear differential equations after expressing
the structure constants in terms of the third derivatives of some
function $\F (a\sp1,\dots,a\sp{N})$
\begin{equation}
\label{cf}
C_{\ ij}^k = \eta^{kl}\F_{ijl}.
\end{equation}
Generically the matrix $\eta$ can be an arbitrary linear
combination with time dependent coefficients
$\eta = \sum_j \alpha\sp{j}({\bf a})\f_j$,
but for simplicity we will mostly consider $\eta=\f_1$.
Then $\eta_{rs}=\iota_\alpha dT_{rs}$.

Various rewritings of the WDVV equations are possible.
In the Frobenius manifold setting there is a pencil of flat connections,
while the general equations (\ref{WDVV}) are equivalent \cite{mor} to
the commuting of matrix-valued vector fields:
$$[\d_i -(\f_j)\sp{-1}\f_i\, \d_j,
   \d_k -(\f_j)\sp{-1}\f_k\, \d_j]=0$$
(with fixed $j$).
For our purposes one may also rewrite (\ref{crossing}) as
\begin{equation}
\label{crossF}
\sum_k C^k_{\ ij}\F_{kln} = \sum_k C^k_{\ il}\F_{kjn}.
\end{equation}
With one form $C=\eta\sp{-1}dT$ (
i.e. $C_{\ ij}^k da\sp{i}= \eta^{kl}\d_i T_{jl}da\sp{i}=
\eta^{kl}\F_{ijl}da\sp{i}$) we have
\begin{equation}
\label{bilinear}
[\C_i, \C_j ]=0\Longleftrightarrow C\wedge C=0
\Longleftrightarrow dT\wedge \eta\sp{-1}dT=0.
\end{equation}
Unfortunately the connection $C$ is unrelated to the various connections
arising in the special K\"ahler geometry because of the (arbitrary)
appearance of $\eta$ which is unrelated to the special K\"ahler metric.
In general the curvature of $C$ is nonvanishing.
(The curvature will, for example, vanish as a consequence of the
WDVV equations if there is a symmetry ${\cal L}_\alpha dT=\mu dT$, for
some matrix $\mu$.)
The final form of (\ref{bilinear}) is reminiscent of the Hirota
bilinear equations which may be interpreted in terms of Pl\"ucker
formulae, but again the appearance of $\eta$ means the bilinear form
in such formulae is not constant.
Whatever, the final form shows that the WDVV equations impose relations
on the period matrices $T$, and consequently restrict the associated
Seiberg-Witten curves that can yield solutions of (\ref{WDVV}).

An important observation was made recently in \cite{BMRWZ} where it was shown
that in the case of dispersionless integrable hierarchies the WDVV
equations maybe derived directly by differentiating corresponding Hirota
algebraic relations. In particular, the solutions associated with
the Landau-Ginzburg topological models can be constructed in this way.
Below, we will reverse the arrow of this implication
and demonstrate that the WDVV equations, if rewritten in terms of second
derivatives, can be simply viewed as certain Hirota-like
functional relations.

The dispersionless Hirota relations (see \cite{BMRWZ} and references
therein about the details and their explicit form) for the second
derivatives, which can be written as
\begin{equation}
\label{hir}
\F_{ij} = T_{ij}({\Bf\varphi})
\end{equation}
where $T_{ij}$ are some {\em known} functions and $\{\varphi_i\}$ denote some
restricted set (of cardinality $N$, the size of the matrices under
consideration in the finite-dimensional situation) second
derivatives, for example
\begin{equation}
\label{vf}
\varphi_i = \F_{1i}({\bf a})
\end{equation}
Without specifying form of the functions $T_{ij}$ in (\ref{hir}) this is
just a {\em dimensional} statement that the {\em matrix} of second
derivatives of any function can be expressed in terms of only a {\em
vector} of variables.

Consider the ``period'' matrix (\ref{pematr})
which for any integrable system is a function of only $N$ variables.
This means that in the ``Siegel upper half-space'' $\{ T_{ij}\}$ of dimension
$\dfrac{N(N+1)}{2}$ we have a ``submanifold'' of dimension $N$
\begin{equation}
\label{hs1}
T_{ij} = T_{ij}(a\sp1,\dots,a\sp{N})
\end{equation}
or codimension $\dfrac{N(N-1)}{2}$ given by defining equations
\begin{equation}
\label{hs2}
f_A(T_{ij})=0
\quad
A=1,\dots,{N(N-1)\over 2}
\end{equation}
Of course, on this submanifold (under a generic assumption of
nonsingularity) one may choose new variables, say
\begin{equation}
\label{defphi}
\varphi_i = T_{1i}({\bf a})
\end{equation}
Consequently there exist functional relations
\begin{equation}
T_{ij} = T_{ij}(\varphi_1,\dots,\varphi_N) = T_{ij}({\bf a}
(\varphi_1,\dots,\varphi_N))
\end{equation}
between the matrix elements of the whole period matrix,
exactly as one has in dispersionless Hirota's relations.
In the case of (\ref{defphi}) all of the matrix elements of
the period matrix would be expressed in terms of a single
row or column. (We give an example of this in section~\ref{ss:eg}.)

More generally, let us assume we can find coordinates $\varphi_k$ so that
\begin{equation}
\label{defc}
dT_{ij}=C^k_{\ ij}\,d\varphi_k.
\end{equation}
Using the definition (\ref{cf}) this means
\begin{equation}
\label{defp}
d\varphi_k=\eta_{kl}d a\sp{l}.
\end{equation}
Both (\ref{defc}) and (\ref{defp}) have the common integrability condition
\begin{equation}
\label{integ}
d\eta_{kl}\wedge d a\sp{l}=0\Longleftrightarrow M\sp{i}_s \F_{irk}=
M\sp{i}_r \F_{isk} ,\qquad M\sp{i}_r=\dfrac{\d \alpha\sp{i}}{\d a\sp{r}},
\end{equation}
where the $\alpha\sp{i}$'s were the coefficients defining $\eta$ above.
These are certainly satisfied for constant $\alpha\sp{i}$. Thus the
choice (\ref{defphi}) would correspond to $\eta=\f_1$.
Supposing the integrability condition (\ref{integ}) holds, our assumption
(\ref{defc}) means that the structure constants can be {\em integrated}, i.e.
defined as derivatives of the functions (\ref{hir})
\begin{equation}
\label{c}
C_{\ ij}^k = {\d T_{ij}\over\d\varphi_k}
\end{equation}
and that
\begin{equation}
\label{inthir}
T_{ij}({\Bf\varphi}) = \int \sum_k C_{ij}^k d\varphi_k,
\end{equation}
i.e. the structure constants determine the form of the Hirota-like
relations (\ref{hir}).

\section{The associativity equations as cocycle conditions}

We shall now derive some consequences of (\ref{defc}).
For a fixed index $a$
$$
d\varphi_k = dT_{ai}\,(\C_a\sp{-1})\sp{i}_{\ k}
$$
whence for any $b$
$$
d\varphi_k = dT_{ai}\,(\C_a\sp{-1})\sp{i}_{\ k}
= dT_{bj}\,(\C_b\sp{-1})\sp{j}_{\ k}.
$$
From these it follows that
\begin{equation}
\label{tcct}
dT_{ai} = dT_{bj} \left(\C_b^{-1}\cdot \C_a\right)^j_{\ i}
= dT_{bj}\left(\C_j^{-1}\cdot \C_i\right)^b_{\ a},
\end{equation}
where the last equality follows from the symmetry of the period matrices.
For fixed $a$ and $b$ we may view these as giving a change of variables
\begin{equation}
T_{ai} = T_{ai}(T_{bj}),\qquad\forall\ i , j,
\end{equation}
and
\begin{equation}
\label{tt}
{\d T_{ai}\over\d T_{bj}} = \left(\C_b^{-1}\cdot \C_a\right)^j_{\ i}
=\left(\C_j^{-1}\cdot \C_i\right)^b_{\ a}.
\end{equation}
These equations may simply be viewed as the chain-rule
\begin{equation}
\label{cv}
{\d T_{ai}\over\d T_{bj}} = \sum_s {\d T_{ai}\over\d \varphi_s}
{\d \varphi_s\over\d T_{bj}} = \left(\C_b^{-1}\cdot \C_a\right)^j_{\ i},
\end{equation}
where we understand (for fixed $b$)
${\d \varphi_s\over\d T_{bj}} = \left(\C_b^{-1}\right)^j_{\ s}$
in terms of the change of variables $\varphi_s\leftrightarrow T_{bj}$.
For future comparison we also record the identity (for fixed $a$, $b$, $c$)
\begin{equation}
\label{3cvid}
\dfrac{\d T_{ai}}{\d T_{bj}}\dfrac{\d T_{ck}}{\d T_{ai}}
\dfrac{\d T_{bl}}{\d T_{ck}}
= (\C_b\sp{-1} \C_a)\sp{j}_{\ i} (\C_a\sp{-1} \C_c)\sp{i} _{\ k}
(\C_c\sp{-1} \C_b)\sp{k} _{\ l}
=(\C_b\sp{-1} \C_a \C_a\sp{-1} \C_c \C_c\sp{-1} \C_b)\sp{j} _{\ l}
=  \delta\sp{j} _l.
\end{equation}

Thus far we have only used change of variables and the fact that the
period matrix (\ref{pematr}) can be expressed in terms of only
genus $g=N$ number of variables.
Such will hold for {\em any} integrable system, but this does not mean
the associativity equations (\ref{WDVV}) or (\ref{crossmatr}) must hold.
Lets now see the meaning of these. From (\ref{tt}) we have
\begin{equation}
\label{3cocy}
\dfrac{\d T_{ai}}{\d T_{bj}}\dfrac{\d T_{bk}}{\d T_{ci}}
\dfrac{\d T_{cl}}{\d T_{ak}}
= (\C_b\sp{-1} \C_a)\sp{j}_{\ i} (\C_c\sp{-1} \C_b)\sp{i} _{\ k}
(\C_a\sp{-1} \C_c)\sp{k} _{\ l}
=(\C_b\sp{-1} \C_a \C_c \sp{-1} \C_b \C_a\sp{-1} \C_c)\sp{j} _{\ l}.
\end{equation}
Now from (\ref{crossmatr}) it follows also that
the matrices of structure constants satisfy (for all $a$ and $b$)
\begin{equation}
\label{crossinv}
\C_a\cdot \C_b^{-1} = \C_b^{-1}\cdot \C_a.
\end{equation}
Substituting (\ref{crossinv}) into (\ref{3cocy}) we conclude that
\begin{equation}
\label{cocycle}
\sum_{i,k}\
\dfrac{\d T_{ai}}{\d T_{bj}}\dfrac{\d T_{bk}}{\d T_{ci}}
\dfrac{\d T_{cl}}{\d T_{ak}}
= \delta\sp{j} _l.
\end{equation}
Equally from (\ref{cocycle}) we deduce (\ref{WDVV}).
Thus the WDVV equations are equivalent to the Hirota-like functional
relations (\ref{cocycle}).
The relation (\ref{cocycle}) is our main statement.

\section{Example: perturbative Seiberg-Witten prepotentials \label{ss:eg}}

Here we shall consider one of the simplest examples of solutions to the WDVV
equations coming from Seiberg-Witten theory. These are related
to $SU(N+1)$ perturbative prepotentials \cite{MMM} and the corresponding
Riemann surface is degenerate. With $a_{ij}\equiv a_i-a_j$ the
perturbative prepotential is given by
\begin{equation}
\label{pertpre}
\F = \2\sum_{i<j}^N a_{ij}^2\log {a_{ij}\over\Lambda} +
\2\sum_{i=1}^N a_i^2\log {a_i\over\Lambda}.
\end{equation}
The second derivatives (\ref{pematr}) (for a special choice of
$\Lambda$) are
\begin{equation}\label{pemape}
\begin{split}
 T_{ij} &= - \log a_{ij}, \ \ i<j; \qquad T_{ij} =
T_{ji},\ \ i>j;\\  T_{ii} &= \ln a_i+\sum_{i<j} \ln a_{ij}
+\sum_{j<i}\ln a_{ji} = \ln a_i - \sum_{j\ne i} T_{ij}.
\end{split}\end{equation}
Using (\ref{pemape}) one may reexpress all matrix elements $T_{ij}$ through
a given row explicitly.
For example, in the first nontrivial case corresponding to $SU(4)$ one finds
\begin{equation}
\label{pemasu4}
\begin{split}
T &= \left(
\begin{array}{ccc}
  x & y & z \\
  y & \log\left(e^{x+y+z}-e^{-y}\right)\left(e^{-z}-e^{-y}\right)-y &
-\log\left(e^{-z}-e^{-y}\right) \\
  z & -\log\left(e^{-z}-e^{-y}\right) &
\log\left(e^{x+y+z}-e^{-z}\right)\left(e^{-z}-e^{-y}\right)-z
\end{array}\right)
\\
\\
&= \left(
\begin{array}{ccc}
 \log\left(e^{\tilde{x}+\tilde{y}+\tilde{z}}+e^{-\tilde x}\right)
\left(e^{-\tilde x}+e^{-\tilde z}\right)- \tilde{x}
& \tilde{x} & -\log\left(e^{-\tilde x}+e^{-\tilde z}\right) \\
  \tilde{x} & \tilde{y} & \tilde{z} \\
-\log\left(e^{-\tilde x}+e^{-\tilde z}\right)  & \tilde{z} &
\log\left(e^{\tilde{x}+\tilde{y}+\tilde{z}}+e^{-\tilde z}\right)
\left(e^{-\tilde x}+e^{-\tilde z}\right)- \tilde{z}
\end{array}\right)
\\
\\
&= \left(
\begin{array}{ccc}
 \log\left(e^{\hat{x}+\hat{y}+\hat{z}}+e^{-\hat x}\right)
\left(e^{-\hat x}-e^{-\hat y}\right)- \hat{x}
& -\log\left(e^{-\hat x}-e^{-\hat y}\right) & \hat{x} \\
-\log\left(e^{-\hat x}-e^{-\hat y}\right)  &
\log\left(e^{\hat{x}+\hat{y}+\hat{z}}+e^{-\hat y}\right)
\left(e^{-\hat x}-e^{-\hat y}\right)- \hat{y} &
\hat{y} \\
\hat{x} & \hat{y} & \hat{z}
\end{array}\right).
\end{split}
\end{equation}
Here we have exhibited the different dependence on the rows
which are taken as independent variables:
$(x,y,z)=(T_{11},T_{12},T_{13})$,
$(\tilde{x},\tilde{y},\tilde{z})=(T_{21},T_{22},T_{23})$ and
$(\hat{x},\hat{y},\hat{z})=(T_{31},T_{32},T_{33})$.

In the $SU(4)$ perturbative Seiberg-Witten case equation (\ref{cocycle})
has essentially only the one nontrivial relation
\begin{equation}
\label{cocsu4}
\sum_{i,k}\
\dfrac{\d T_{2i}}{\d T_{1j}}\dfrac{\d T_{1k}}{\d T_{3i}}
\dfrac{\d T_{3l}}{\d T_{2k}}
= \delta\sp{j} _l.
\end{equation}
The corresponding matrices may be straightforwardly computed
using (\ref{pemape}). It is easy to check, that (\ref{cocsu4}) holds
provided the Hirota relations (cf. with \cite{ZaHi}) are satisfied:
\begin{equation}
\label{h1}
\sign(j-i)e^{-T_{ij}} + \sign(k-j)e^{-T_{jk}} + \sign(i-k)e^{-T_{ki}} = 0,
\qquad
i \neq j \neq k,
\end{equation}
(originating from $a_{ij} + {\rm cyclic}=0$) together with
\begin{equation}
\label{h2}
e^{T_{11}+T_{12}+T_{13}-T_{23}} + e^{T_{13}+T_{23}+T_{33}-T_{12}} -
e^{T_{12}+T_{22}+T_{23}-T_{13}} = 0,
\end{equation}
(coming from $a_ia_{jk} + {\rm cyclic}=0$) and
\begin{equation}
\label{h3} e^{T_{12} + T_{22}+T_{23}} - e^{T_{13} + T_{33} +
T_{23}} = e^{-T_{23}}
\end{equation}
(realizing $a_{ij}=a_i-a_j$). These may be seen to be satisfied
upon utilising (\ref{pemape}).

\section{Discussion}

Although the WDVV equations arise in several different physical settings
no unifying geometry as yet underpins them. With the additional restrictions
of topological field theory Frobenius manifolds successfully encode
the geometry, but the restrictions are too severe to allow other interesting
examples coming from Seiberg-Witten theory and the dispersionless limits of
solutions to the Hirota equations.
In this note we have pursued this link between the WDVV equations and
the algebraic relations arising from the dispersionless Hirota
equations \cite{BMRWZ}. The key idea was to focus on the second derivatives
of the prepotential and the connections between them.
First we noted a rather general phenomenon, independent of the
associativity equations. Any ``generalized period matrix'' (\ref{pematr})
of an integrable system implies the existence of certain subspace
of matrix elements of the period matrices.
As such, there are certain functional relations between the matrix elements.
When we additionally impose the associativity or WDVV equations
we obtained a set of equivalent functional relations (\ref{cocycle}).
From this perspective we can easily understand why not every Seiberg-Witten
curve, or the spectral curve of every integrable system, will lead to
solutions of the WDVV equations: only very special subspaces lead to
solutions. The WDVV impose bilinear relations (\ref{bilinear})
on the differentials of the period matrix restricted to this subspace.

Whereas in the case of dispersionless hierarchies the Hirota relations
(\ref{hir}) have a rather simple form (they are algebraic, or the
functions $T_{ij}(\Bf\varphi)$ are polynomials) in the general setting
this is not the case,
and that is why we prefer to call them {\em functional} Hirota relations.
Our functional relations (\ref{cocycle}) depend on a choice of
three different indices $a$, $b$ and $c$.
These three cycles are similar to the three points usually chosen when
one writes down the conventional Hirota equations
(see, for example \cite{ZaHi}).
In the setting where $T_{ij}$ plays the role of the period matrix of
a Riemann surface (of genus $g$), this choice of $a$, $b$ and $c$
corresponds to a choice of three different {\em cycles} on the corresponding
Riemann surface. Here the Schottky relations of the period matrices
reduce the dimension of $g(g+1)/2$ symmetric matrices to a space of
dimension $3g-3$; the constraint coming from integrability reduces this
still more (to in general $g$), with further restriction from the
WDVV-functional relations.
Our final example illustrated the functional relations for a
perturbative Seiberg-Witten solution.

\section*{Acknowledgements}

We are grateful to E.F.~Corrigan and A.~Zabrodin for useful discussions.
This research was supported by the NATO grant R81478. A.M. thanks the
hospitality at University of Edinburgh, where the work was completed.
The work of A.M. was also partially supported by by RFBR grant
No.~01-01-00539, INTAS grant No.~00-561
and the grant for the support of scientific schools
No.~01-02-30024.


\end{document}